\documentclass[aps,prl,twocolumn,showpacs,superscriptaddress,groupedaddress]{revtex4}

\usepackage{amsmath,amsfonts}
\usepackage{bm}
\usepackage{graphicx}
\usepackage[pdftex,usenames,dvipsnames]{color}
\usepackage{comment}

\def\opone{\leavevmode\hbox{\small1\kern-3.8pt\normalsize1}}

\usepackage{subfigure}

\begin{document}
\title{Properties of a rare-earth-ion-doped waveguide at sub-Kelvin temperatures for quantum signal processing}
\author{N. Sinclair}
\affiliation{Institute for Quantum Science and Technology, and Department of Physics \& Astronomy, University of Calgary, Calgary, Alberta T2N 1N4, Canada}
\author{D. Oblak}
\affiliation{Institute for Quantum Science and Technology, and Department of Physics \& Astronomy, University of Calgary, Calgary, Alberta T2N 1N4, Canada}
\author{C. W. Thiel}
\affiliation{Department of Physics, Montana State University, Bozeman, Montana 59717, USA}
\author{R. L. Cone}
\affiliation{Department of Physics, Montana State University, Bozeman, Montana 59717, USA}
\author{W. Tittel}
\affiliation{Institute for Quantum Science and Technology, and Department of Physics \& Astronomy, University of Calgary, Calgary, Alberta T2N 1N4, Canada}

\begin{abstract}
We characterize the 795 nm $^3$H$_6$ to $^3$H$_4$ transition of Tm$^{3+}$ in a Ti$^{4+}$:LiNbO$_{3}$ waveguide at temperatures as low as 800 mK. Coherence and hyperfine population lifetimes -- up to 117 $\mu$s and 2.5 hours, respectively -- exceed those at 3 K at least ten-fold, and are equivalent to those observed in a bulk Tm$^{3+}$:LiNbO$_{3}$ crystal under similar conditions. We also find a transition dipole moment that is equivalent to that of the bulk. Finally, we prepare a 0.5 GHz-bandwidth atomic frequency comb of finesse $>$2 on a vanishing background. These results demonstrate the suitability of rare-earth-doped waveguides created using industry-standard Ti-indiffusion in LiNbO$_3$ for on-chip quantum applications.
\end{abstract}
\maketitle

Integrated optics offers the possibility of scalable manipulation of light due to small guiding volume, chip size, ever-improving fabrication methods, and low loss \cite{sohler2008, bazzan2015}. These attractive features have also enabled advanced experiments and applications of quantum optics and quantum information processing, with many demonstrations using telecommunication-industry-standard Ti-indiffused LiNbO$_3$ waveguides, e.g. see \cite{saglamyurek2011, tanzilli2012, saglamyurek2014, sinclair2014, bonneau2012, carolan2015, meany2015} and references therein. Moreover, the ability to combine many components onto a single substrate is required for the implementation of an integrated quantum information processing node that performs local operations, interconnects, and measurements for quantum-secured computing and communications \cite{kimble2008}. For example, a chip containing multiplexed sources of entangled photon pairs, Bell-state analyzers, photon number-resolving nondestructive photon detection, and feed-forward qubit-mode translation and selection, e.g. using frequency shifts or on-demand quantum memories, could pave the way to the development of a practical quantum repeater \cite{sangouard2011, sinclair2014, sinclair2015}. A promising avenue to this end, and to quantum information processing in general, is based on qubit interfaces using rare-earth-ion-doped (REI-doped) crystals at cryogenic temperatures \cite{saglamyurek2011, tittel2010, saglamyurek2014, sinclair2015, hedges2010, bussieres2013, deriedmatten2015}. Consequently, the development of rare-earth-ion-doped waveguides constitutes an exciting and important path towards applications.

Many ground-breaking quantum optics experiments, in particular quantum memory for light \cite{hedges2010, tittel2010, bussieres2013, deriedmatten2015}, are based on cryogenically-cooled REIs doped into a variety of bulk crystals. However, work employing REI-doped crystalline waveguides has, until only the past year, been restricted to LiNbO$_3$ -- generally doped with thulium -- into which waveguides were fabricated by means of titanium indiffusion \cite{staudt2007, saglamyurek2011, saglamyurek2014, sinclair2014, sinclair2015}. A likely explanation for the lack of investigations using other REIs in Ti$^{4+}$:LiNbO$_3$ is that an earlier low-temperature characterization of Tm$^{3+}$:Ti$^{4+}$:LiNbO$_{3}$ at 3 K revealed non-ideal properties for quantum signal processing \cite{sinclair2010}. Specifically, optical coherence and hyperfine-level lifetimes were shown to be significantly reduced compared to those of a bulk Tm$^{3+}$:LiNbO$_{3}$ crystal at the same temperature \cite{sun2012}, possibly due to perturbations stemming from the introduction of Ti. As alternatives, several groups have recently explored the use of Y$_2$SiO$_5$ crystals together with evanescent wave-mediated planar waveguides \cite{marzban2015}, ion beam-milled photonic crystals \cite{zhong2015}, or waveguides created via femtosecond laser writing \cite{corrielli2015} -- all of which guide light in unperturbed regions of the REI-doped crystal. While the results are promising, it is still unclear how the waveguide fabrication process affects the relevant REI properties for quantum applications and if good properties can be retained using industry-standard Ti-indiffusion in LiNbO$_3$.

To investigate this question, we measure optical properties of the $^3$H$_6$ to $^3$H$_4$ transition of a Tm$^{3+}$:Ti$^{4+}$:LiNbO$_{3}$ waveguide using photon echoes and spectral hole burning \cite{macfarlane1987} at temperatures as low as 800 mK and with applied magnetic fields up to 600 Gauss. We find at least one order of magnitude improvement of critical properties (optical coherence and hyperfine population lifetimes) compared to those measured previously in a similar waveguide at 3 K \cite{sinclair2010}. Moreover, the measured properties match those observed using Tm$^{3+}$:LiNbO$_{3}$ bulk crystals under similar conditions \cite{thiel2016}, and improve upon those observed with other REI-doped waveguides \cite{marzban2015,zhong2015,corrielli2015}. In addition, our measurements establish a ratio of coherence lifetime to ($^3$H$_4$) excited-state population lifetime that is greater than one \cite{sinclair2010, sinclair2016}, which benefits applications in the field of cavity quantum electrodynamics \cite{mcauslan2009} or employing cross-phase modulation \cite{sinclair2015}. To show key requirements for efficient quantum information processing, we burn persistent holes to transparency, and tailor a 0.5 GHz-bandwidth atomic frequency comb (AFC) \cite{bussieres2013, deriedmatten2015} with a finesse of three on a vanishing absorption background using magnetic fields up to 20 kG. Finally, we measure Rabi frequencies that indicate a bulk-equivalent \cite{sun2012} transition dipole moment of $\sim$10$^{-32}$ C$\cdot$m. This establishes that waveguide fabrication does not reduce the dipole moment -- an important finding in view of light-matter coupling.

\textit{Experimental methods--}Measurements are carried out using a 15.7 mm-long  Ti$^{4+}$:Tm$^{3+}$:LiNbO$_{3}$ waveguide. It is created by increasing the index of refraction of a \mbox{$\sim$4 $\mu$m-wide} strip through indiffusion of Ti into the surface of a 0.9 mm-thick, 0.7\% Tm-indiffusion-doped lithium niobate crystal \cite{sinclair2010}. The crystal is then mounted inside a pulse tube cryocooler on a Cu stage attached to a GGG salt pellet that generates temperatures as low as 800 mK (measured near the crystal) via adiabatic demagnetization. Light is directed into, and out of, the waveguide using single-mode fiber butt-coupling. Transmission through the entire cryogenic setup is 10-20\%, mainly limited by imperfect overlap between fiber and waveguide modes as well as reflections from uncoated surfaces. Magnetic fields of up to 20 kG are applied parallel to the crystal's c-axis using a superconducting solenoid coil. We employ an external-cavity diode laser, producing polarized light set normal to the crystal's c-axis \cite{sinclair2010, sun2012}. Its wavelength is tuned between 795.5 and 799.0 nm (vac.) using a diffraction grating. These wavelengths are chosen to ensure the optical depth of the crystal is no more than approximately 1.4 and below a value where optical spectroscopy becomes difficult. A 400 MHz acousto-optic modulator is employed to produce pulses as short as 50 ns for photon echo measurements, or as long as several ms for hole burning and spectral tailoring. Spectral features are probed by ramping the laser frequency using a LiNbO$_3$ waveguide phase modulator driven with a sawtooth-modulated voltage \cite{saglamyurek2014, sinclair2014}. Optical transmission is detected using a 1 GHz (2 MHz) AC (DC)-coupled amplified diode, digitized by an oscilloscope. To minimize errors due to laser power fluctuations, jitter, and noise we average over many repetitions of the same experiment. Furthermore, in cases where it is undesired, persistent spectral hole burning is mitigated by slowly sweeping the laser frequency.

\begin{figure}
\begin{center}
\includegraphics[width=\columnwidth]{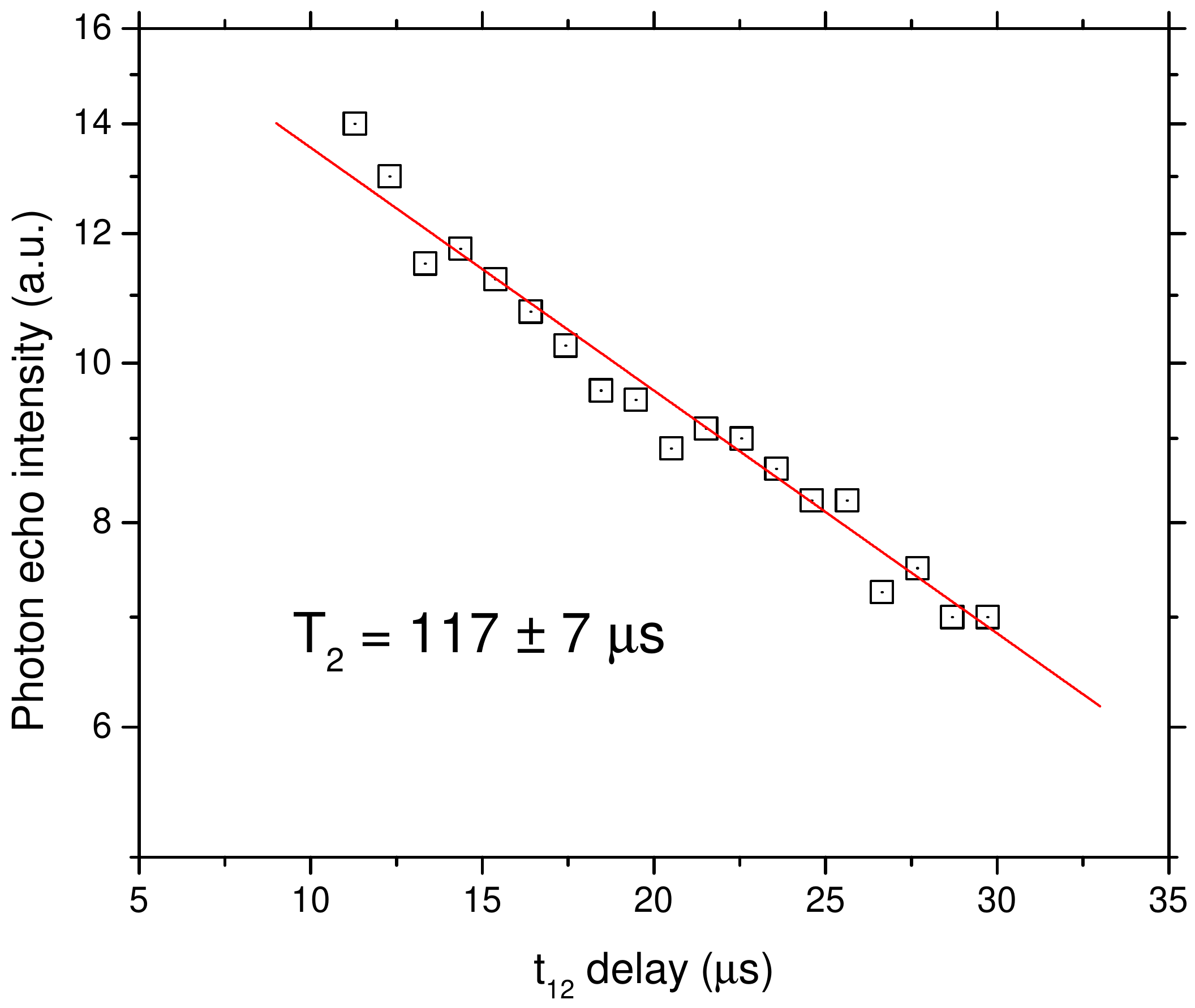}
\caption{Two pulse photon echo decay. Measurements at a wavelength of 795.6 nm, using a 300 G magnetic field, and at a temperature of 810 mK yield a coherence lifetime of \mbox{117 $\mu$s}.} 
 \label{fig:t12decay}
\end{center}
\end{figure}

\textit{Results--}Long optical coherence lifetimes are a key reason why rare-earth-ion-doped crystals are employed for quantum applications \cite{tittel2010, bussieres2013, deriedmatten2015}. By observing a two-pulse photon echo decay, we measure the coherence lifetime $T_2$ of the transition at 795.6 nm in a 300 G magnetic field, at a temperature of 810 mK, and with a laser peak power inside the waveguide of $\sim$0.4 mW (set to reduce excitation-induced decoherence (EID) \cite{thiel2014} while simultaneously ensuring the echo intensity is above the noise level). The decay, as shown in Fig. \ref{fig:t12decay}, is fitted using $I(t_{12}) = I_0 \textrm{exp}[{-2(\tfrac{2 t_{12}}{T_2} )^x}]$,
where $t_{12}$ is the delay between the two excitation pulses, $I_0$ is the echo intensity at $t_{12}=0$, and $x$ describes the decay shape \cite{mims1968}. The decay shape is exponential and does not exhibit any time-dependent increase of decoherence, i.e. $x=1$. We find a coherence lifetime of 117 $\mu$s; 70 times greater than what we measured in Tm$^{3+}$:Ti$^{4+}$:LiNbO$_{3}$ at 3.5 K \cite{sinclair2010}, and larger than the 86 $\mu$s we observed using a Tm$^{3+}$:LiNbO$_{3}$ bulk crystal at 790 mK temperature, 794.25 nm wavelength, and $\sim$150 Gauss magnetic field \cite{thiel2016}. The increase of $T_2$ with lower temperatures is due to a reduction of phonon-induced decoherence. We expect that $T_2$ can be increased further by probing ions that absorb at wavelengths towards the center of the inhomogeneous line, as these ions are more characteristic of typical crystal environments than the ions currently probed. We note that the larger $T_2$ of the waveguide in comparison to that measured in the corresponding bulk crystal under similar conditions was not expected. This difference may be due to additional EID during the bulk crystal $T_2$ measurement, where the intensity of the excitation pulses was a factor of 5 larger than those employed here, or possibly elevated temperatures due to poor thermal contact between the crystal and its mount. However, more experiments are needed to verify these explanations.

\begin{figure}
\begin{center}
\includegraphics[width=\columnwidth]{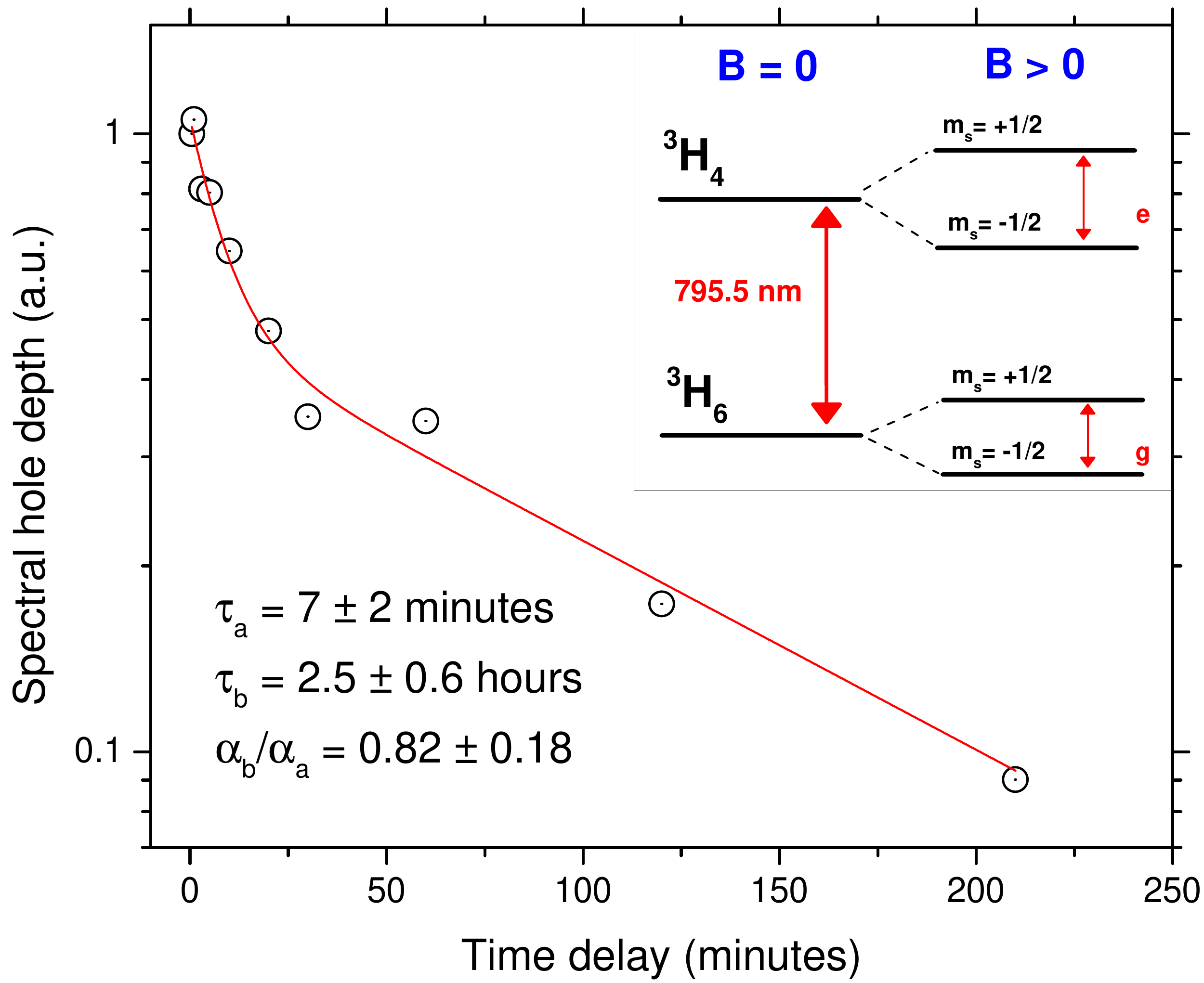}
\caption{Persistent spectral hole burning decay. Measurements at a wavelength of 795.5 nm, under a magnetic field of \mbox{600 G}, and at a temperature of 850 mK yield a population decay that follows a double exponential. Inset: Simplified energy level diagram (not to scale) showing optically excited levels and hyperfine levels split by the applied magnetic field.} 
 \label{fig:pshb}
\end{center}
\end{figure}

In conjunction with a sufficiently-long $T_2$, many quantum applications require the ability to spectrally tailor the absorption profile. This requires long-lived sub-levels to optically pump ions, e.g. to create transparency or efficient memories. Therefore, we investigate the dynamics of the Tm hyperfine sub-levels under 600 G field, at a temperature of 850 mK, and at a wavelength of 795.5 nm using time-resolved persistent spectral hole burning. In an external magnetic field, the hyperfine sub-levels are non-degenerate due to an enhanced effective nuclear Zeeman effect from the hyperfine coupling of the $^{169}$Tm spin-1/2 nucleus with the electronic states \cite{macfarlane1987}. This splits both the excited and ground levels into pairs ($m_s = \pm 1/2$) with a difference in energy splitting of $|\Delta_e-\Delta_g|\sim$140 kHz/G \cite{sun2012, sinclair2016}, allowing optical pumping of the nuclear spin populations between hyperfine levels. The resultant spectral hole in the optical absorption decays as the nuclear spin population returns to thermal equilibrium as is shown in Fig. \ref{fig:pshb}, with a simplified sub-level structure depicted in the inset. 
We fit the decay using a double exponential, $\Delta \alpha (t_d) = \Delta \alpha_a e^{-t_d/{\tau_a}}+ \Delta \alpha_b e^{-t_d/{\tau_b}}$, where $t_d$ is the time delay between hole burning and probing, $\tau_a$ and $\tau_b$ are two different hole lifetimes, and $\Delta\alpha_a$ and $\Delta\alpha_b$ are the relative amplitudes of the two population components present in the hole decay. We find two exponentials, similar to what we measured in this material at 3.5 K \cite{sinclair2010}, except now with lifetimes of 7 minutes, and 2.5 hours, respectively. These lifetimes are two orders of magnitude greater than those measured at 3.5 K \cite{sinclair2010}, and agree well with those measured using a Tm$^{3+}$:LiNbO$_3$ bulk crystal under similar conditions \cite{thiel2016}. Indeed, reduction in phonon scattering at lower temperatures predicts an increase in population lifetimes of hyperfine levels. Although we have not yet investigated the origin of the double decay, we conjecture that two components may arise from inequivalent Tm$^{3+}$ sites in the lattice or possibly from a combination of optical pumping of the lithium and niobium spins through the superhyperfine interaction, in addition to pumping of the Tm$^{3+}$ nuclear spin states \cite{sun2012,sinclair2016}. The double decay at this temperature seems to be intrinsic to the material and not a result of waveguide fabrication since a similar behavior is also observed in the bulk crystal \cite{thiel2016}.

\begin{figure}
\begin{center}
\includegraphics[width=\columnwidth]{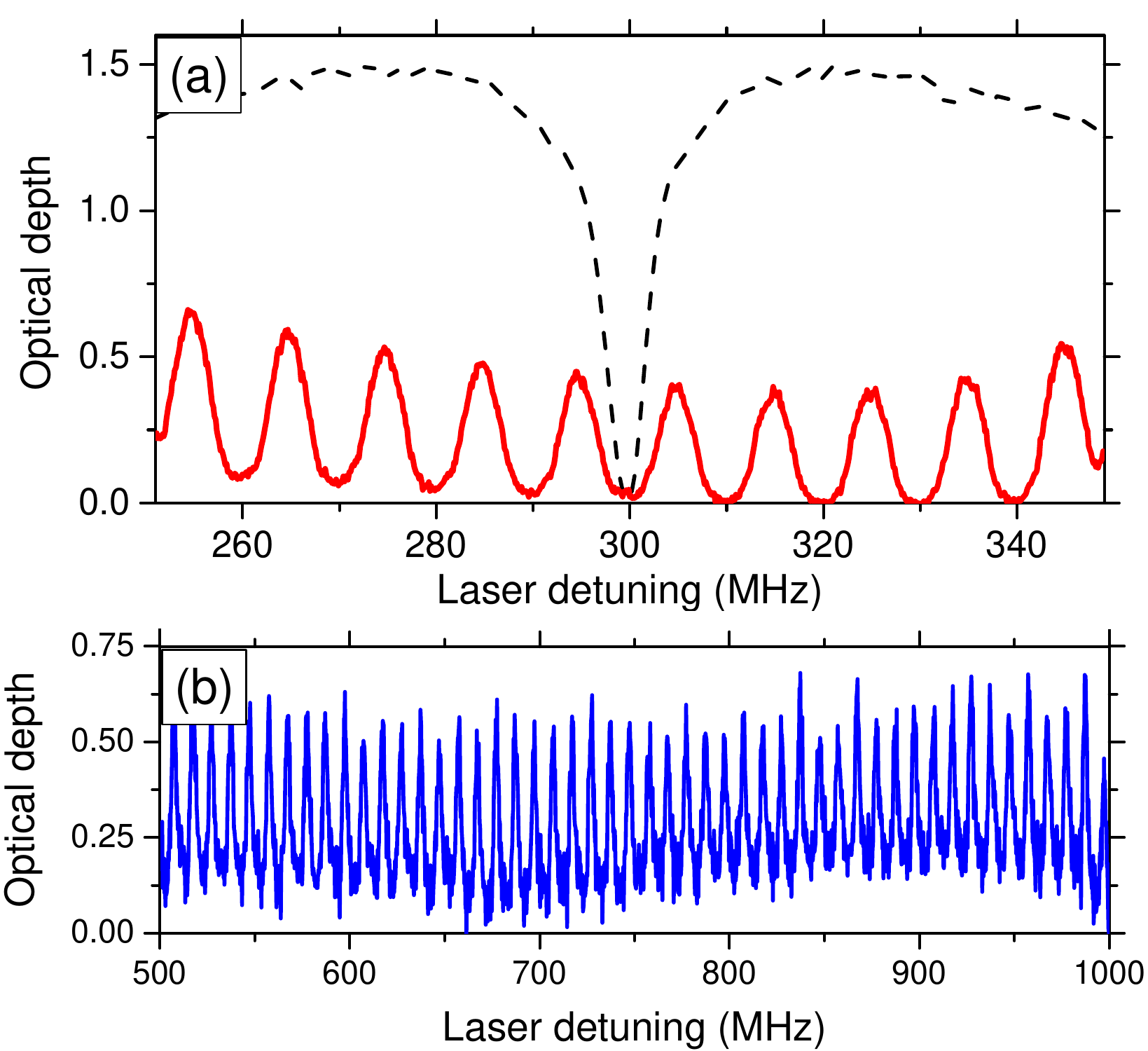}
\caption{(a) A spectral hole (dashed curve) and a 100 MHz-broad AFC (solid curve) created under 500 G and 2 kG magnetic fields, respectively, at a temperature of 800 mK and a wavelength of 795.5 nm. (b) A 0.5 GHz AFC prepared on near-zero background.} 
 \label{fig:holeafc}
\end{center}
\end{figure}

We also demonstrate that our Tm-doped waveguide may be used for efficient quantum applications that require the preparation of an arbitrary (and broadband) persistent spectral feature on a vanishing absorption background \cite{tittel2010}. The largest bandwidth of such a feature for Tm-doped materials is determined by $|\Delta_e-\Delta_g|$ to ensure that ions are optically-pumped to hyperfine levels that are detuned from the feature. First, improving upon what we have shown at 3 K and what has been shown with other rare-earth-ion-doped waveguides, we generate a persistent spectral hole with a vanishing absorption background. Specifically, at a temperature of 800 mK, wavelength of 795.5 nm, and using a 500 G magnetic field, we burn a spectral hole during several ms, wait for 100 ms, and then, with reduced laser power, sweep the laser frequency over the hole to measure its absorption profile. The laser sweep is repeated several times and then the laser wavelength is tuned off-resonance to 799 nm to determine the intensity level at full transparency. The result of this measurement is shown by the dashed curve in Fig. \ref{fig:holeafc}a.
The $\sim$5 MHz width of the spectral hole is attributed to power broadening, laser jitter, and the resolution of the laser sweep (each of these contributions is verified by independent measurements). Note that we have also observed holes with jitter-limited widths of 200 kHz, using low excitation powers and short excitation durations (to reduce the impact of power broadening and laser jitter, respectively), but not to transparency (results shown in the Supplemental Material). Although we have not yet quantified the effect of time-dependent spectral diffusion over 100 ms timescales under similar conditions, we believe that spectral diffusion does not play a role as initial measurements have shown no linewidth broadening over timescales up to $\sim$1 ms at 300 G magnetic field and a temperature of 800 mK. Observations using a Tm$^{3+}$:LiNbO$_3$ bulk crystal at 1.7 K and with 138 G field have also shown no time-dependent spectral diffusion \cite{sun2012, sinclair2016}.

Next we increase the magnetic field to 2 kG and then perform the same hole burning sequence at different frequency detunings to create multiple spectral holes, each separated by 10 MHz from its neighbor. The result is a 100 MHz-bandwidth AFC programmed with a 100 ns storage time and finesse $F$, defined by the product of storage time and the linewidth of each tooth comprising the comb, of approximately two. Our AFC is shown by the solid curve in Fig. \ref{fig:holeafc}a. The residual background at the AFC edges is due to a range of energy splittings between the hyperfine levels, i.e. a hyperfine inhomogeneous broadening, that causes $|\Delta_e-\Delta_g|$ to take a distribution of frequencies, and leads to ions being pumped into hyperfine levels that are not sufficiently detuned from the AFC. We note that the hyperfine broadening increases at a rate of $\sim$50 kHz/G at this wavelength \cite{sinclair2016}. The broadening rate is consistent with observations using a Tm$^{3+}$:LiNbO$_3$ bulk crystal \cite{sun2012} and is expected to decrease significantly at wavelengths closer to 794.3 nm \cite{sun2012}. The reduced optical depth of the AFC compared to that of the single hole is due to power broadening and excitation during the read sequence. Note that the 100 ms wait time is $>$30 times greater than that used for storage of non-classical light \cite{saglamyurek2011}, and chosen to demonstrate AFC persistence. 

To show that our waveguide is suitable for high data-rate applications employing AFCs with $F$ beyond 2, we increase the field to 20 kG, and repeat the experiment, except now preparing a 0.5 GHz-bandwidth AFC programmed with 100 ns storage time and $F=3$. The result is shown in Fig. \ref{fig:holeafc}b (a 100 MHz-bandwidth section of this AFC is shown in the Supplemental Material). Note that the efficiency of this AFC is 0.4\%, leading to a total system efficiency of around 0.1\%. As before, the small optical depth is caused by the read sequence, while the small residual background is due to laser intensity modulation from optical cavities formed by waveguide and fiber endfaces that leads to a detuning-dependent pumping and reading efficiency. We expect that excitation at wavelengths closer to 794.3 nm will reduce the bandwidth-limiting impact of hyperfine inhomogeneous broadening and allow creating AFC bandwidths of several GHz with vanishing absorption background using fields of tens of kilogauss \cite{sun2012}.

\begin{figure}
\includegraphics[width=\columnwidth]{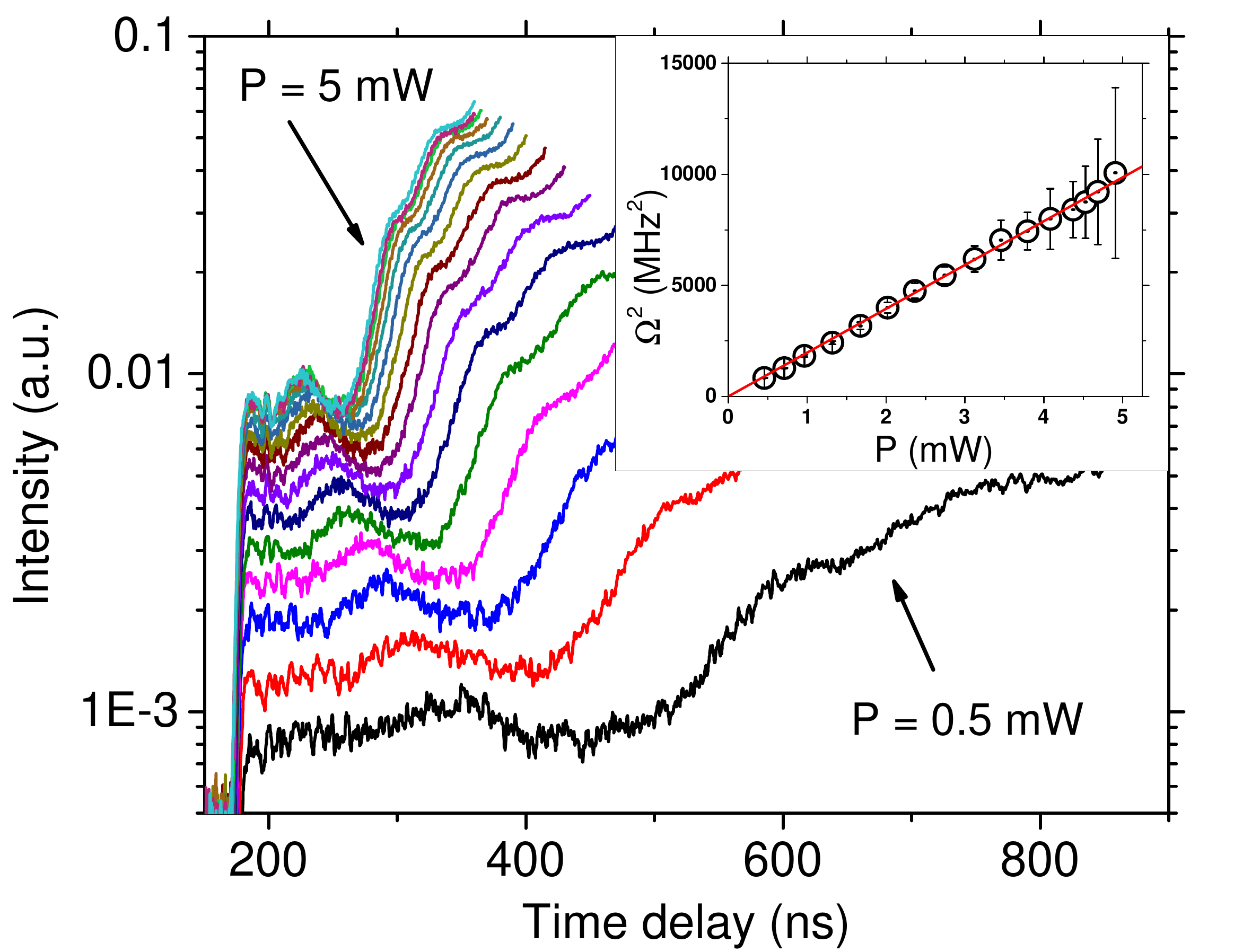}
\caption{Measurements of the transition dipole moment. Observed optical nutation (log scale) using pulses of power $P$ varied between 0.5 and 5 mW (shown with different colors). Curves are truncated to timescales where nutation is most visible. Inset: Power dependence of $\Omega^2$. We calculate an average $\Omega$ from each nutation using the first inversion maximum and minimum, and the second and third inversion maxima.} 
 \label{fig:dipole}
\end{figure}

Due to confinement of light, a waveguide allows for higher-frequency Rabi oscillations per photon than using a bulk optics arrangement. Thus, waveguides are desirable for both on- and off-resonant light-matter interaction, e.g. spectral tailoring, Raman interaction, or cross-phase modulation. To these ends, we quantify the magnitude of the $^3$H$_6$ to $^3$H$_4$ transition dipole moment $\mu$ of our waveguide at a temperature of 800 mK under zero magnetic field using both optical nutation (at a wavelength of 795.3 nm) and two-pulse photon echo excitation (at a wavelength of 795.6 nm). The strength of the light-matter coupling is scaled by a ratio of $\mu^2 /A$ per photon, where $A$ is the mode cross-sectional area. We note it is crucial that the magnitude of $\mu$ be preserved after waveguide fabrication to obtain the advantage from having a smaller $A$. To measure $\mu$, we direct a pulse into the waveguide and detect a modulation of the pulse intensity (i.e. nutation) after it exits the waveguide -- see Fig. \ref{fig:dipole}.
The time-dependent decay of the nutation indicates a distribution of Rabi frequencies due to transverse (Gaussian intensity distribution) and longitudinal (propagation loss) variation in the laser intensity as well as different atomic transitions (inhomogeneous broadening). The modulation frequency is related to the Rabi frequency $\Omega$ experienced by the ions at the center of the Gaussian beam and at the front of the waveguide (i.e. at peak intensity). Using the method described in Ref. \cite{sun2000}, we find a linear $\Omega^2$-dependence, see the inset of Fig. \ref{fig:dipole}, which we fit using $\Omega^2 = \kappa \frac{\mu^2}{A}  P$. Here, $P$ is the estimated power in the waveguide, $\kappa ={2 (n^2+2)^2}/{(9 n c \epsilon_0 \hbar^2)}$ is a material constant \cite{thiel2014}, $n$ is the index of refraction, and $A\approx \pi$ (6.25 $\mu$m)$^2$ is estimated from independent beam-waist measurements. We find $\mu = (3.7 \pm 0.4) \times 10^{-32}$  C$\cdot$m, agreeing well with $\mu = 4 \times 10^{-32}$  C$\cdot$m reported for a Tm$^{3+}$:LiNbO$_{3}$ bulk crystal \cite{sun2012}. This dipole moment is one of the largest of all rare-earth-ion optical transitions that have been studied \cite{macfarlane1987,mcauslan2009}. To confirm our result, we perform a series of two-pulse echo excitation measurements (with results shown in the Supplemental Material), finding $\mu = (7.3 \pm 0.1) \times 10^{-32}$ C$\cdot$m, which is comparable to that extracted from our more precise nutation measurements.

\textit{Conclusion}--At 800 mK we measure dramatically improved spectroscopic properties of Tm$^{3+}$:Ti$^{4+}$:LiNbO$_{3}$ compared to previous work at 3 K \cite{sinclair2010}, with performance surpassing that of other rare-earth-ion-doped waveguides \cite{marzban2015,zhong2015,corrielli2015}, and with properties of the ions within the Ti-modified host matrix matching those in the corresponding bulk Tm$^{3+}$:LiNbO$_3$ crystal \cite{sun2012, thiel2016}. Our results advance the development of low-loss quantum light-matter integrated circuits using industry-standard materials and suggest that crystal modification may not affect important properties of rare-earth ions. However, more detailed investigations of, e.g., field-dependent spectral diffusion beyond 1 ms time delays and wavelength-dependent characteristics are needed to ascertain this conjecture.

\begin{acknowledgments}
We thank E. Saglamyurek and C. Deshmukh for discussions, and M. George, R. Ricken, and W. Sohler for fabrication of the waveguide. We acknowledge funding through Alberta Innovates Technology Futures (AITF), the Natural Sciences and Engineering Research Council of Canada (NSERC), the Defense Advanced Research Projects Agency (DARPA) Quiness program (Contract No. W31P4Q-13-1-0004), the US National Science Foundation (NSF) under award nos. PHY-1415628 and CHE-1416454, and the Montana Research and Economic Development Initiative. Furthermore, W.T. acknowledges support as a Senior Fellow of the Canadian Institute for Advanced Research (CIFAR).
\end{acknowledgments}

\end{document}